\documentclass[lettersize,journal]{IEEEtran}
\usepackage{amsmath,amsfonts}
\usepackage{algorithmic}
\usepackage{algorithm}
\usepackage{array}
\usepackage[caption=false,font=normalsize,labelfont=sf,textfont=sf]{subfig}
\usepackage{textcomp}
\usepackage{stfloats}
\usepackage{url}
\usepackage{verbatim}
\usepackage{graphicx}
\graphicspath{{./fig/}}
\usepackage{cite}
\begin{document}

\title{Deep Brain Ultrasound Ablation Thermal Dose Modeling with \textit{in Vivo} Experimental Validation}

\author{Zhanyue Zhao, Benjamin Szewczyk, Matthew Tarasek, Charles Bales, Yang Wang, Ming Liu, Yiwei Jiang, Chitresh Bhushan, Eric Fiveland, Zahabiya Campwala, Rachel Trowbridge, Phillip M. Johansen, Zachary Olmsted, Goutam Ghoshal, Tamas Heffter, Katie Gandomi, Farid Tavakkolmoghaddam, Christopher Nycz, Erin Jeannotte, Shweta Mane, Julia Nalwalk, E. Clif Burdette, Jiang Qian, Desmond Yeo, Julie Pilitsis, and Gregory S. Fischer

\thanks{Z. Zhao, B. Szewczyk, C. Bales, Y. Wang, M. Liu, Y. Jiang, K. Gandome, F. Tavakkolmoghaddam, C. Nycz, and G. S. Fischer are with Worcester Polytechnic Institute, Worcester, MA e-mail: zzhao4@wpi.edu, gfischer@wpi.edu.}

\thanks{B. Szewczyk, J. Qian, and J. Pilitsis are with the Department of Neurosurgery, Albany Medical Center, Albany, NY}

\thanks{M. Tarasek, C. Bhushan, E. Fiveland, and D. Yeo are with GE Global Research Center, Niskayuna, NY}

\thanks{Z. Campwala, R. Trowbridge, Z. Olmsted, S. Mane, J. Nalwalk, J. Qian, and J. Pilitsis are with the Department of Neuroscience and Experimental Therapeutics, Albany Medical Center, Albany, NY}

\thanks{P. M. Johansen and J. Pilitsis are with Charles E. Schmidt College of Medicine, Florida Atlantic University, Boca Raton, FL}

\thanks{E. Jeannotte is with Animal Resources Facility, Albany Medical Center, Albany, NY}

\thanks{G. Ghoshal, T. Heffter, and E. C. Burdette are with Acoustic MedSystems, Inc., Savoy, IL}

\thanks{This research is supported by National Institute of Health (NIH) under the National Cancer Institute (NCI) under Grant R01CA166379 and R01EB030539.}
}

\markboth{Journal of \LaTeX\ Class Files,~Vol.~14, No.~8, August~2021}%
{Shell \MakeLowercase{\textit{et al.}}: A Sample Article Using IEEEtran.cls for IEEE Journals}

\maketitle

\begin{abstract}

Intracorporeal needle-based therapeutic ultrasound (NBTU) is a minimally invasive option for intervening in malignant brain tumors, commonly used in thermal ablation procedures. This technique is suitable for both primary and metastatic cancers, utilizing a high-frequency alternating electric field (up to 10 MHz) to excite a piezoelectric transducer. The resulting rapid deformation of the transducer produces an acoustic wave that propagates through tissue, leading to localized high-temperature heating at the target tumor site and inducing rapid cell death. To optimize the design of NBTU transducers for thermal dose delivery during treatment, numerical modeling of the acoustic pressure field generated by the deforming piezoelectric transducer is frequently employed. The bioheat transfer process generated by the input pressure field is used to track the thermal propagation of the applicator over time. Magnetic resonance thermal imaging (MRTI) can be used to experimentally validate these models. Validation results using MRTI demonstrated the feasibility of this model, showing a consistent thermal propagation pattern. However, a thermal damage isodose map is more advantageous for evaluating therapeutic efficacy. To achieve a more accurate simulation based on the actual brain tissue environment, a new finite element method (FEM) simulation with enhanced damage evaluation capabilities was conducted. The results showed that the highest temperature and ablated volume differed between experimental and simulation results by 2.1884$^\circ$C (3.71\%) and 0.0631 cm$^3$ (5.74\%), respectively. The lowest Pearson correlation coefficient (PCC) for peak temperature was 0.7117, and the lowest Dice coefficient for the ablated area was 0.7021, indicating a good agreement in accuracy between simulation and experiment.

\end{abstract}


\section{Introduction}
Thermal ablation techniques are being developed as minimally invasive or non-invasive alternatives to primary and metastatic brain tumor surgery \cite{mcdannold2010transcranial}. Different types of technologies were developed, including four conventional methods: (1) Radio frequency (RF) ablation, which uses electricity from an alternating current at a favorite frequency of around 500kHz to produce ionic agitation in tissue, which results in friction and heating of tumor \cite{ni2005review,decadt2004radiofrequency}. (2) Microwave ablation (MWA), which uses electromagnetic waves up to 2450MHz to generate polar molecules agitation with friction and heat in tissue and causes coagulative necrosis of cells \cite{simon2005microwave}. (3) Laser ablation, which utilizes intense and highly-collimated beams of monochromatic light that are designed for biological tissues for thermal destruction of tumors \cite{mahamood2018laser,renk2012basics}. (4) Ultrasound ablation, which uses the transmission of high-frequency sound waves through biological tissue transfers mechanical energy that leads to localized heating and coagulative necrosis of surrounding cells \cite{kennedy2005high}. The ultrasound ablation contains two types of techniques, namely extracorporeal methods using high intensity focused ultrasound (HIFU) for non-invasive ablation procedure \cite{zhou2011high}, and intracorporeal method using an inserted transducer with direct contact of the target area for minimally invasive ablation procedure \cite{diederich1996transurethral,diederich2004transurethral}. There is a promising new ablation technology developed in recent years called pulsed-field ablation (PFA), which uses a train of microsecond duration high amplitude electrical pulsed that ablates myocardium by electroporation of the sarcolemmal membrane without measurable tissue heating, and this method is aiming for the treatment of atrial fibrillation\cite{bradley2020pulsed,cochet2021pulsed}.

Numerical modeling of thermal ablation solves the physical equations governing energy deposition and heat transfer in tissue to determine the transient temperature profile and assess tissue damage upon heating \cite{prakash2012considerations}. However, due to patient-specific tissue compositions that might affect local features like absorption, attenuation, and perfusion, thermal ablation can be challenging to anticipate with absolute accuracy. Finite element modeling (FEM) can be applied in thermal ablation modeling. It can calculate the piezoelectric transducer's acoustic pressure and the transferring bioheat deposition along the desired acoustic medium by subdividing a large system into smaller ones. It can also simpler elements by meshing the construction. The finite element method formulation of a boundary value problem finally results in a system of algebraic equations. 

In this work, an extended FEM simulation towards NBTU application in a real brain tissue environment with tissue damage evaluation is one of the primary studies for MR-guided robot-assisted conformal ablation procedures. We developed a Comsol FEM simulation based on a real brain tissue medium with a CEM43 isodose map for evaluating tissue damage. Moreover, we validated the model with \emph{in vivo} swine ablation experimental results and proved the feasibility and accuracy of this new model.

\section{FEM Analysis of NBTU in Brain Tissue}
Temperature measurements can provide absolute or relative temperature changes in comparison to unheated tissue or reference data\cite{rieke2007referenceless}. However, thermal damage is generally evaluated using the Sapareto-Dewey equation, and the ablation zone is defined by the cumulative equivalent minute (CEM) standard\cite{elias2013magnetic, sapareto1984thermal}. The CEM43 is defined as the equivalent time the target tissue temperature has to be at 43$^\circ$C to induce thermal damage\cite{elias2013magnetic}. We selected a CEM43 of 70 based on our collaborators' experience with ablation in liver tissue\cite{misra2015trimodal, misra2015bi, ghoshal2013situ},  also a comparison to similar paper\cite{n2014active} and studies demonstrating a CEM43 of 50 to 240 range to induce cell death with soft tissue\cite{damianou1994effect}. Using Sapareto and Dewey's formulation, the thermal damage dosage was computed using the temperature from the MRTI given by Equation \ref{equ:CEM43}

\begin{equation}
    CEM43=\sum_{t=0}^{t=final} R^{43-T_t}\Delta t, \left\{
        \begin{aligned}
            & R=0.25\ for\ T<43^\circ C\\
            & R=0.50\ for\ T\ge43^\circ C
        \end{aligned}
    \right.
    \label{equ:CEM43}
\end{equation}

where $T_t$ is the average temperature during time $\Delta t$. The unit of thermal dose is equivalent to minutes at 43$^\circ$C. The application of CEM43 can estimate the damage to brain tissue.

Based on the thermal damage of soft tissue, a FEM simulation can be created and used as a numerical prediction tool for accurately tracking thermal changes using NBTU in preparation for human trials.

\subsection{FEM Analysis of NBTU in Brain Tissue}

The FEM study for simulating thermal propagation and CEM43 threshold calculation was informed by NBTU parameters during surgery, while the parameters for the application were based on the phantom study \cite{gandomi2019thermo, gandomi3d}. We enhanced our existing numerical models with an acoustic medium based on brain-tissue-specific parameters for comparison to our \emph{in vivo} animal data. 

\subsubsection{Simulation Setup}

\begin{figure}
    \centering
    \includegraphics[width=2.5in]{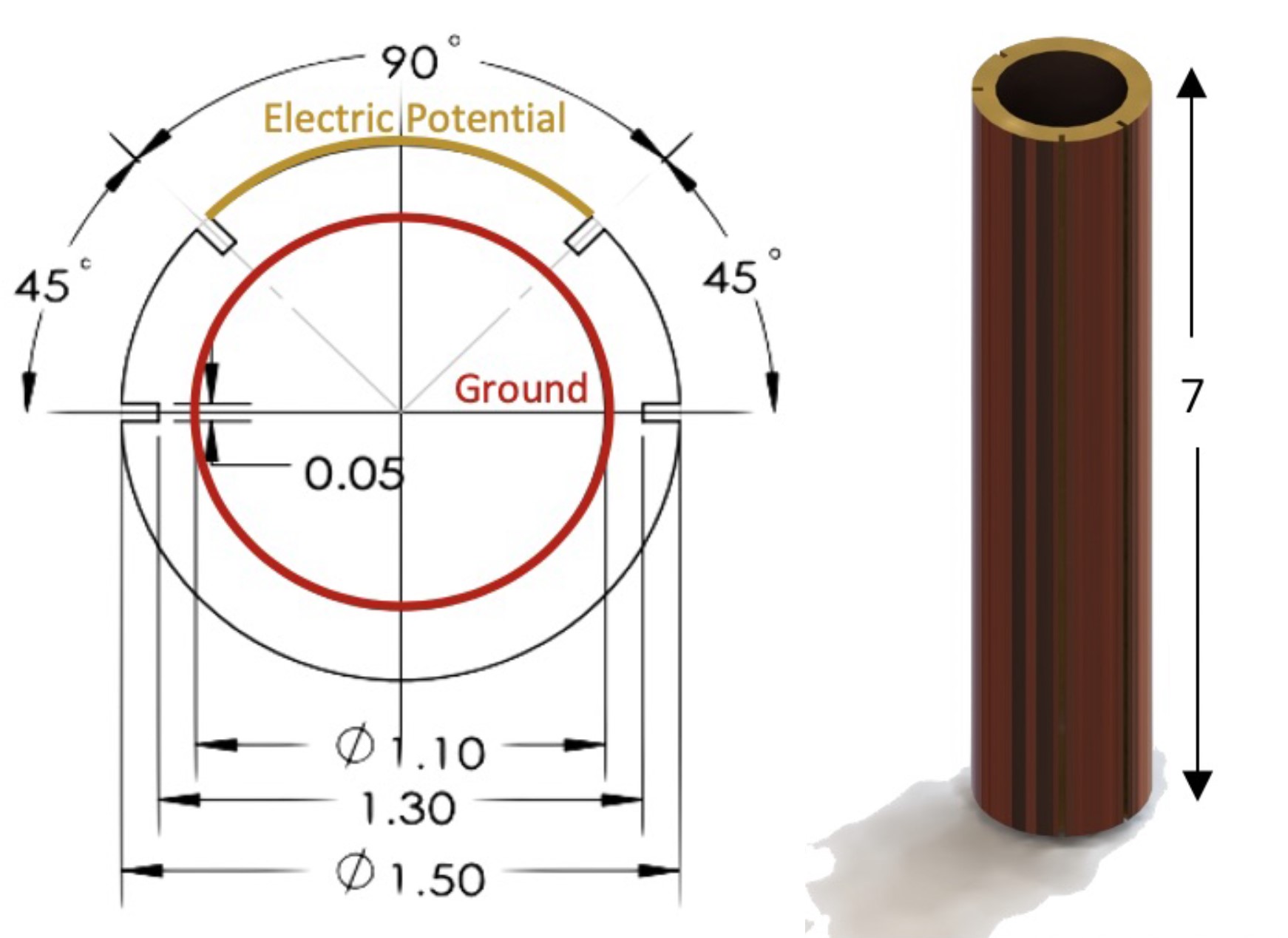}
    \caption{Schematic of the PZT-4 transducer used by the applicator is also shown with units in mm. Image reproduced from \cite{gandomi2020modeling}.}
    \label{fig:geo}
\end{figure}

The FEM model was implemented in Comsol to model the probe’s mechanical deformation, resultant applicator stationary acoustic pressure field, and the thermal damage isodose threshold of 70 CEM43. A cross-sectional model of the probe’s geometry was created with a dimensional parameter of 1.5mm outer diameter by 1.1mm inner diameter ring, and our notches of 0.1mm depth by 0.05mm width were also added such that the probe was segmented into 90$^\circ$ and 180$^\circ$ sectors. A detailed drawing of the probe dimension is shown in Fig. \ref{fig:geo}. Lead zirconate titanate (PZT-4) from Comsol material library was selected as the probe cylinder material with no further modifications and the simulation properties for the transducer were used as shown in Table \ref{tbl:NBTUproper}. The probe was surrounded by a 100 mm $\times$ 100 mm (L$\times$W) 2D acoustic medium. 

\begin{table}[]
    \centering
    \caption{Simulation properties of the modeled NBTU applicator.}
    \label{tbl:NBTUproper}
    \begin{tabular}{|l|l|}
    \hline
    \multicolumn{2}{|c|}{Transducer Properties}           \\\hline
    Ceramic Material   & PZT-4                          \\\hline
    Density            & 7500 ($kg/m^3$) \\\hline
    Resonant Frequency & 5.0444 MHz                     \\\hline
    \end{tabular}
\end{table}

The solid mechanics physics interface defined the transducer as a piezoelectric material and defined a fixed constraint within the inner surface of the cylinder. The piezoelectric poling direction was defined by a base vector coordinate system defined by the transform shown in Table \ref{tbl:vector} and was aligned radially outward from the center of the transducer geometry. The electrostatics physics interface was also used to define the electric potential, or electrical terminal along the 90$^\circ$ outer surface of the transducer and to define a ground constraint along the inner surface. Different types of probes (ablation range) were used in the experiment, and the specific parameters of probes that were used in the simulation are shown in Table \ref{tbl:3probepara}. The electric potential can be calculated by Equation \ref{equ:potential}:

\begin{equation}
    V = \sqrt{\frac{P(1+\varepsilon)Z}{\eta}}
    \label{equ:potential}
\end{equation}

Where $P$ is the desired acoustic power we need for ablation, $\varepsilon$ is the power loss generated from the RF inline filter, and the power loss for these probes has a range of approximately 10-20\% based on the probe type and the method of fabrication, based our collaborator's suggestion we used median value of 15\% in our simulation. $Z$ is the transducer impedance value, and we used 50ohm as a general piezoelectric material value. $\eta$ is the efficiency of the probe, detailed value can be found in Table \ref{tbl:3probepara}.

\begin{table}[]
    \centering
    \caption{The base vector coordinate frame is used to give the poling direction of the piezoelectric element.}
    \label{tbl:vector}
    \begin{tabular}{|c|c|c|c|}
    \hline 
    Base Vectors & x                 & y                 & z \\\hline 
    x1           & -sin(atan2(Y, X)) & -cos(atan2(Y, X)) & 0 \\\hline 
    x2           & 0                 & 0                 & 1 \\\hline 
    x3           & cos(atan2(Y, X))  & sin(atan2(Y, X))  & 0 \\\hline 
    \end{tabular}
\end{table}

\begin{table*}[]
    \centering
    \caption{Simulation properties of 3 types of probes, and physical parameters were used.}
    \label{tbl:3probepara}
    {
    \begin{tabular}{|c|c|c|c|c|c|}
    \hline
    \begin{tabular}[c]{@{}c@{}}Probe\\ Number\end{tabular} & \begin{tabular}[c]{@{}c@{}}Degree of\\ Ablation\end{tabular} & \begin{tabular}[c]{@{}c@{}}Physical\\ Frequency\\ (MHz)\end{tabular} & \begin{tabular}[c]{@{}c@{}}Efficiency\\ $\eta$ \end{tabular} & \begin{tabular}[c]{@{}c@{}}Simulation\\ Eigenfrequency\\ (MHz)\end{tabular} & \begin{tabular}[c]{@{}c@{}}Electric\\ Potential\\ (V)\end{tabular} \\ \hline
    \begin{tabular}[c]{@{}c@{}}BRN.360.1\\ 0.00028-R\end{tabular} & 360$^\circ$ & 6.837 & 69\% & 5.0555 & \begin{tabular}[c]{@{}c@{}}14.7442 (3W)\\ 17.0251 (4W)\end{tabular} \\ \hline
    \begin{tabular}[c]{@{}c@{}}BRN.180.1\\ 0.00029-R\end{tabular} & 180$^\circ$ & 6.900 & 53\% & 5.1772 & \begin{tabular}[c]{@{}c@{}}16.8232 (3W)\\ 19.4257 (4W)\end{tabular} \\ \hline
    \begin{tabular}[c]{@{}c@{}}BRN.90.180.2\\ 0.00030-R\end{tabular} & 90$^\circ$ & 7.157 & 55\% & 5.1778 & 23.3550 (6W) \\ \hline
    \end{tabular}
    }
\end{table*}

The pressure acoustics physics interface was used to define the wave propagation through the medium as a linear elastic fluid model with an attenuation coefficient of 31.96 Np/m. The bioheat transfer physics interface was used to define a heat source applied to the medium with a user-defined activation function shown in equation 1, where \emph{Q} is the absorbed ultrasound energy over the acoustic field, $acpr$. $Q_{pw}$ is the dissipated power intensity, \emph{step} is a step function from 0 to 1 with smoothing of 0.005, t is the current time step of the simulation, and $t_{ProbeOn}$ is a constant that represents the total time the probe was activated.

The brain tissue referenced acoustic medium material properties are described in Table \ref{tbl:brainmedium}. Far-field and thermally insulated boundary conditions were applied to the acoustic medium, and no reflection of ultrasonic waves from edges was assumed. A frequency domain study was conducted to obtain the acoustic pressure field produced by the applicator, applying a tetrahydral mesh with a maximum element size of $\lambda/6$, where $\lambda$ represented the wavelength of the produced ultrasonic wave within the medium.

\begin{table}[]
    \centering
    \caption{Material properties of brain tissue based acoustic medium. Data imported from \cite{sadeghi2016parameter,de2016heat,vaupel1989blood}.}
    \label{tbl:brainmedium}
    \begin{tabular}{|ll|}
    \hline
    \multicolumn{2}{|c|}{Acoustic Medium Properties}            \\ \hline
    \multicolumn{1}{|l|}{Heat Capacity ($C$)}        & 3680 ($J/kg/^\circ C$)  \\ \hline
    \multicolumn{1}{|l|}{Density ($\rho$)}              & 1035.5 ($kg/m^3$) \\ \hline
    \multicolumn{1}{|l|}{Thermal Conductivity (\emph{K})} & 0.565 ($W/m/^\circ C$)  \\ \hline
    \multicolumn{1}{|l|}{Blood Density ($\rho_b$)}        & 1050 ($kg/m^3$) \\ \hline
    \multicolumn{1}{|l|}{Blood Perfusion ($\omega_b$)}      & 0.013289 ($kg/m^3/s$) \\ \hline
    \multicolumn{1}{|l|}{Speed of Sound (\emph{c})}       & 1460 ($m/s$)     \\ \hline
    \multicolumn{1}{|l|}{Attenuation ($\alpha_{atten}$)}          & 31.95 Np/m     \\ \hline
    \multicolumn{1}{|l|}{Metabolic Heat ($Q_m$)}       & 16229 ($W/m^3$)   \\ \hline
    \end{tabular}
\end{table}

\subsubsection{Acoustic Pressure Mapping}

\begin{figure}
    \centering
    \includegraphics[width=\columnwidth]{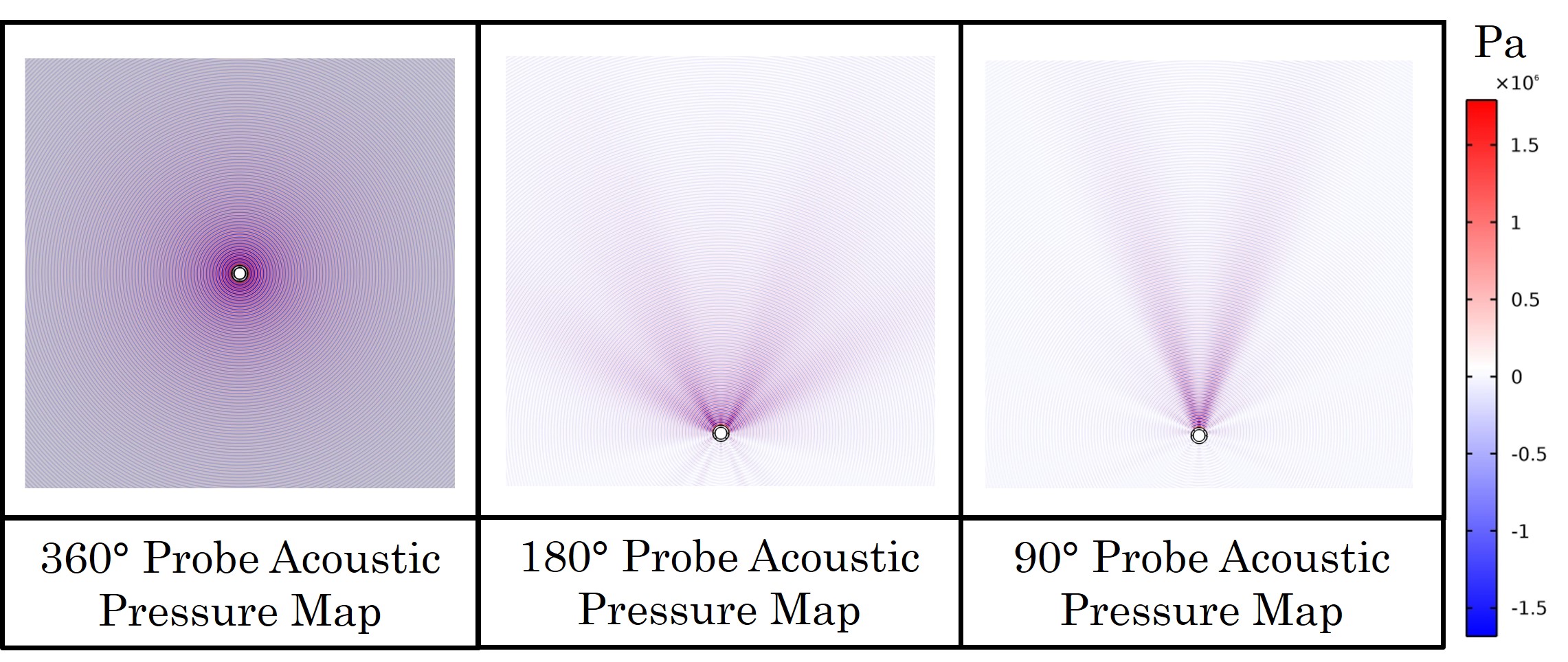}
    \caption{Acoustic pressure field mapping pattern of 3 types of probes.}
    \label{fig:acpr}
\end{figure}

The generated ultrasound waves produced by the transducer can be simulated in Comsol using a frequency domain study at the selected resonant mode, and the pattern mapping can be found in Fig. \ref{fig:acpr}. Considering the properties of brain tissue performed small change in speed of sound propagation in both mediums, the patterns are similar to our previous simulation work reported in \cite{gandomi2020modeling, gandomi2019thermo, gandomi3d, szewczyk2022happens}.

\subsubsection{CEM43 Mapping}
\label{sec:c6s2ss2}

The bioheat transfer time-dependent study was conducted to validate the heat deposition by the simulated acoustic pressure field, and we modified the default equation in this node by using the generalized Penne’s bioheat transfer equation shown in Equation \ref{equ:pennesBHT}:

\begin{equation}
    \rho C\frac{\partial T}{\partial t}=\nabla \cdot k \nabla T-c_b\rho_b\omega_b(T-T_b)+Q_m+Q_p
    \label{equ:pennesBHT}
\end{equation}

Where $\rho$ ($kg/m^3$) is the tissue density, $C$ ($J/kg/^\circ C$) is the tissue specific heat capacity, k ($W/m/^\circ C$) is the tissue thermal conductivity, T ($^\circ C$) is the current temperature, $T_b$ ($^\circ C$) is the temperature of blood, and we assumed the initial temperature of both tissue and blood as 37$^\circ$C. $\rho_b$ ($kg/m^3$) is the blood density, $\omega_b$ ($kg/m^3/s$) is the blood perfusion, $c_b$ ($J/kg/^\circ C$) is the blood specific heat capacity, which we used the same value as the brain tissue heat capacity ($C$). $Q_p$ ($W/m^3$) is the heat deposition due to sonication given by Equation \ref{equ:depositionch}

\begin{equation}
    Q_p=\frac{\alpha_{atten}p^2}{\rho c}
    \label{equ:depositionch}
\end{equation}

Where $\alpha_{atten}$ (Np/m) is the medium attenuation, p ($Pa$) is the acoustic pressure field, $\rho$ ($kg/m^3$) is the medium density, c ($m/s$) is the speed of sound within the medium. $Q_m$ ($W/m^3$) is metabolic heat, which is the heat generated by the tissue. 

\begin{figure}
    \centering
    \includegraphics[width=\columnwidth]{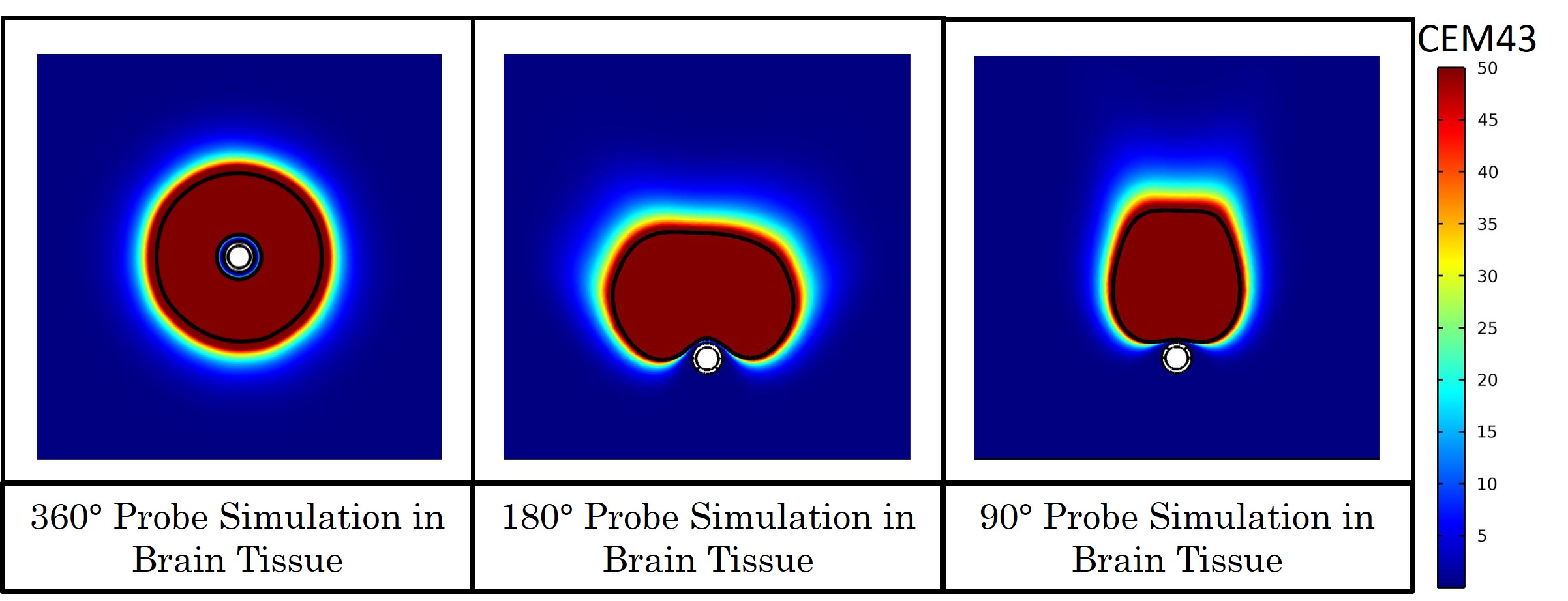}
    \caption{Demonstration of the CEM43 of 70 thermal dose map for 3 types of probes, from left to right namely 4W under 120s, 4w under 180s, and 6w under 690s. The solid black line represents the 70 CEM43 isodose lines.}
    \label{fig:ablationpat}
\end{figure}

A general form PDE study was then conducted to calculate the CEM43 thermal dose map and 70 CEM43 isodose lines, which are shown in Fig. \ref{fig:ablationpat}. Predicted thermal dose maps for NBTU probes showed inhomogeneous CEM 43 thermal dose patterns because of the random blood perfusion and metabolic heat influence generated by Comsol. Using the same volume acquisition method for MRT images, we used surface integration to calculate the 70 CEM43 isodose lines surrounding the area of the transducer in the middle section. Areas of additional slides with 5 mm slice thickness were acquired, multiplied by 5 mm, and summed together to calculate the numerical predicted ablated volume. 

\subsubsection{Model Sensitivity Analysis}

We also performed a Taguchi-based sensitivity analysis with the above model to determine the contribution and ranking of parameters. We considered the maximum obtained temperature as the response variable, and the six 3-level parameters, namely Thermal Conductivity ($k$),  Heat Capacity ($C$), Metabolic Heat ($Q_m$), Blood Perfusion ($\omega_b$), Attenuation ($\alpha_{atten}$), and Voltage (V) applied onto the transducer, as factors of interest. The 3-level of each parameter based on human soft tissue proprieties, namely low, medium, and high, were shown in Table \ref{tbl:sentivitypara}. To obtain the minimum number of simulations needed for the parameters with corresponding levels we used Taguchi analysis standard L-27 orthogonal array. After obtaining all combinations of parameters and their corresponding maximum temperature, we then performed the analysis of variance (ANOVA) to quantify the effect of each factor on the response variable \cite{sudharsan2000parametric, ng2008engineering}. Calculations of Taguchi design and ANOVA were all carried out in Minitab 21 software (Minitab LLC., PA). Results of ANOVA are shown in Table \ref{tbl:ANOVA}. As a result, V majorly affects the maximum obtained temperature with more than 50\% contribution in the model. The second and third major contributing factors are $\alpha_{atten}$ and $k$, respectively. The aforementioned three parameters contribute over 83\% in total to the response variable. $C$, $Q_m$, and $\omega_b$ are not significant in our model.

\begin{table*}[]
\centering
    \caption{Parameters with 3 levels used for sensitivity analysis. Data imported from \cite{sadeghi2016parameter, de2016heat, vaupel1989blood, thermalC, HeatC,jamil2013ranking,  toga2002brain, aminoff1968handbook, bakris2017hypertension, choi2019comparing}.}
    \label{tbl:sentivitypara}
    {
    \begin{tabular}{|l|l|c|c|c|}
    \hline
    Factor                   & Units    & Level 1 (low) & Level 2 (medium) & Level 3 (high) \\ \hline
    Thermal Conductivity ($K$) & (W/m/$^\circ$C)  & 0.18          & 0.39             & 0.56           \\ \hline
    Heat Capacity ($C$)        & (J/kg/$^\circ$C) & 1300          & 3000             & 3800           \\ \hline
    Metabolic Heat ($Q_m$)    & (W/m$^3$)   & 700           & 4200             & 65400          \\ \hline
    Blood Perfusion ($\omega_b$)     & (kg/m$^3$/s)    & 0.0002        & 0.00057          & 0.015          \\ \hline
    Attenuation  ($\alpha_{atten}$)         & (Np/m)     & 20            & 35               & 60             \\ \hline
    Voltage ($V$)              & (V)      & 15            & 25               & 35             \\ \hline
    \end{tabular}
    }
\end{table*}

\begin{table*}[]
\centering
    \caption{ANOVA table for factors of interest.}
    \label{tbl:ANOVA}
    {
    \begin{tabular}{|l|c|c|c|c|c|c|c|}
    \hline
    Factor & DF & ADJ SS & ADJ MS  & F-Value & P-Value & \%Contribution & Ranking \\ \hline
    $K$      & 22 & 22639  & 11319.7 & 13.03   & 0.0001  & 15.436         & 3       \\ \hline
    $C$     & 2  & 1925   & 962.4   & 1.11    & 0.357   & 1.313          & 6       \\ \hline
    $Q_m$     & 2  & 4173   & 2086.7  & 2.40    & 0.127   & 2.845          & 5       \\ \hline
    $\omega_b$     & 2  & 5704   & 2852.0  & 3.28    & 0.068   & 3.889          & 4       \\ \hline
    $\alpha_{atten}$     & 2  & 25361  & 12680.7 & 14.60   & 0.000   & 17.292         & 2       \\ \hline
    V      & 2  & 74700  & 37349.8 & 43.00   & 0.000   & 50.933         & 1       \\ \hline
    Error  & 14 & 12159  & 868.5   &         &         &                &         \\ \hline
    Total  & 26 & 146662 &         &         &         &                &         \\ \hline
    \end{tabular}
    }
\end{table*}

\section{MRTI Imaging Validation}

This section summarized the experiments from the year 2020 to 2022 and selected swine data was used for validation. All the subject was approved by the Albany Medical College (AMC) Institutional Animal Care and Use Committee (IACUC). MR delivery of NBTU was conducted on male and female Sus scrofa domestics swine (8 to 20 weeks old, 10 weeks on average) weighing between 18 and 25kg. One acute swine (sacrificed immediately post-procedure), one subacute swine (sacrificed 1 to 4 days post-procedure), and 5 survival swines were selected to create a total of 7 lesions. MRT-images acquired from a 3T GE Signa Architect scanner (GE Global Research, Niskayuna, NY, USA) and the ablated volume calculated from MRT-images were compared with simulation results \cite{campwala2021predicting}. To operate the interstitial thermal ablation procedure, a seven-DOF NeuroAblation robot was created as an MR-compatible device. The end effector of the NeuroAblation robot is modular and may be changed to accommodate the preferred applicator for the intended intervention. In this work, a needle-based therapeutic ultrasound (NBTU) probe developed by Acoustic MedSystems Inc (Illinois, United States) powered by the TheraVision control system was deployed in the subject. A detailed description can be found in our previous work \cite{szewczyk2022happens, campwala2021predicting, tavakkolmoghaddam2021neuroplan, jiang2024icap}.

\subsection{In-Vivo Swine Ablation Experiment}

In this study, pigs were positioned in a prone position using a custom head holder to align with a NeuroAblation robot, which was fixed to MRI bed rails and registered with the MRI scanner's coordinate system. The procedure involved dividing the MRI room into a sterilized zone for the surgeon and nurse and a standard zone for the robot operator. The entry and target points for the procedure were selected using 3D MRI images, and the NeuroAblation robot validated the coordinates.

Once the points were confirmed within the robot's range, the pig was removed from the scanner, and the robot was operated to reach the target position, except for needle insertion, which was performed by the surgeon. Afterward, the pig was repositioned in the scanner for the ablation procedure. Thermal dose mapping was done using intraoperative MRI to track temperature changes and assess the ablation's effectiveness.

The procedure included low-dose test ablations to confirm the ablation volume's shape and direction before full treatment. Post-ablation imaging was conducted, and pigs were euthanized at various time points for further analysis. The study measured temperature distribution, thermal damage, and ablated volumes to assess the effectiveness of the procedure. A detailed description can be found in \cite{szewczyk2022happens, campwala2021predicting}.

\subsection{MRT-Images Temperature Validation}

\begin{figure*}
    \centering
    \includegraphics[width=1\linewidth]{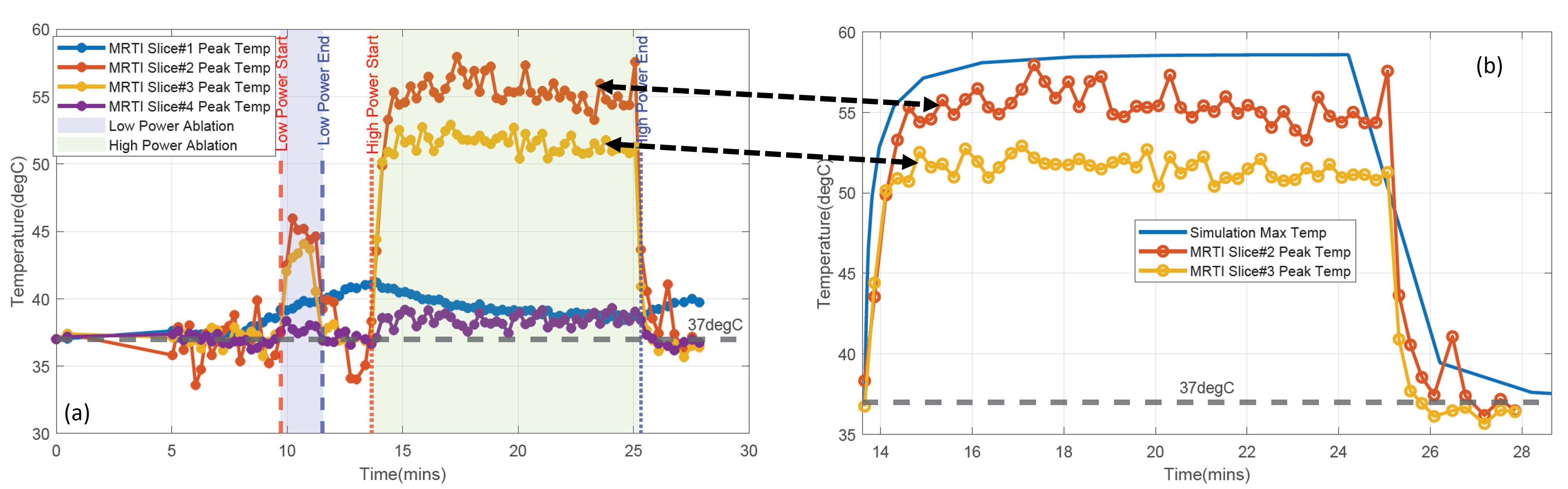}
    \caption{Data from a 90$^\circ$ probe with 6w acoustic power under 690s duration time of ablation. (a) MRTI temperature change versus time with five 5mm-spaced slices. The whole process consisted of a short time range of low-power ablation for pre-testing and a high-power ablation range. (b) Selected slices (slice 2 and 3) versus the simulation result in temperature change over time. The Pearson correlation coefficient between slice 2 data and simulation peak temperature is 0.9527.}
    \label{fig:P90}
\end{figure*}

7 swine experimental data were selected for simulation validation. Firstly, the 90$^\circ$ probe with 6w acoustic power under 690s duration time of ablation (swine 7 experimental and simulation results) was selected as one of the examples of analysis in this section, which consisted of the short period of low power ablation for pre-testing, and a high power ablation range, and the MRTI peak temperature change over time and compare with the simulation maximum temperature results are shown in Fig. \ref{fig:P90}. Note that the ablation process concludes both heating up and cooling down duration because during the cooling down period, the surrounding area was still being heating up, and the heating continued contributing to the CEM43 thermal isodose map. Results indicate that the center ablation surface performed between slice 2 and 3 considering only the two slices changed temperature significantly while the other slices changed very small, and the maximum temperature increased up to 57$^\circ$C. The right figure shows the selected slices (slice 2 and 3) versus the simulation result in temperature change over time. Based on the simulation the predicted maximum temperature reached up to 58.5823$^\circ$C, which yields maximum peak temperature deference with 1.5823$^\circ$C (2.78\%). The Pearson correlation coefficient (PCC) metrics were evaluated to compare the experimental and simulation results. The PCC between experimental data from slice 2 and the simulation peak temperature curve reached 0.9527.

\begin{figure*}
    \centering
    \includegraphics[width=0.81\linewidth]{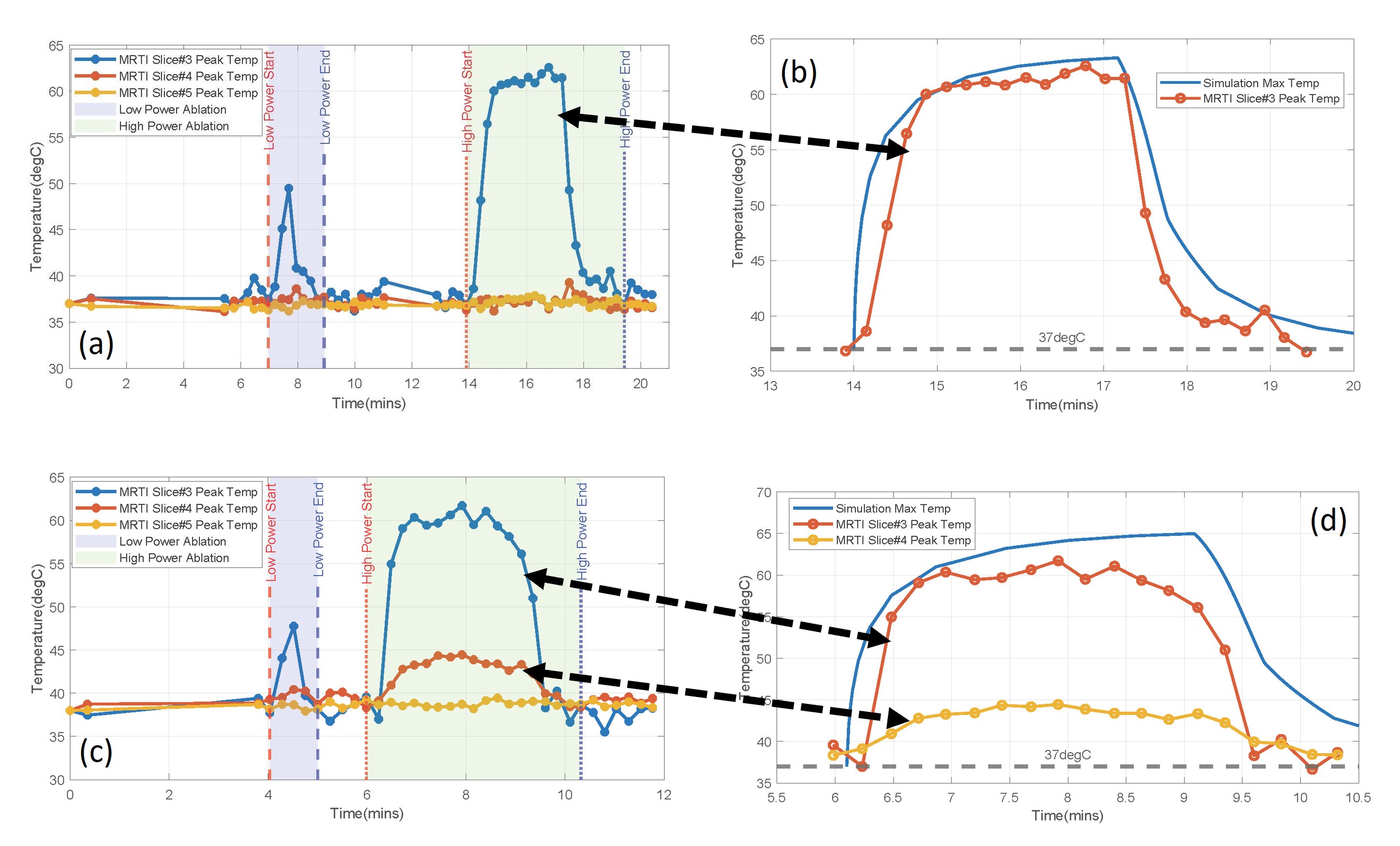}
    \caption{Multiple MRTI results versus the simulation data using 180$^\circ$ probe. (a)-(b) Swine 4 experimental and simulation results. (c)-(d) Swine 6 data. }
    \label{fig:MRT180}
\end{figure*}

\begin{figure*}
    \centering
    \includegraphics[width=0.81\linewidth]{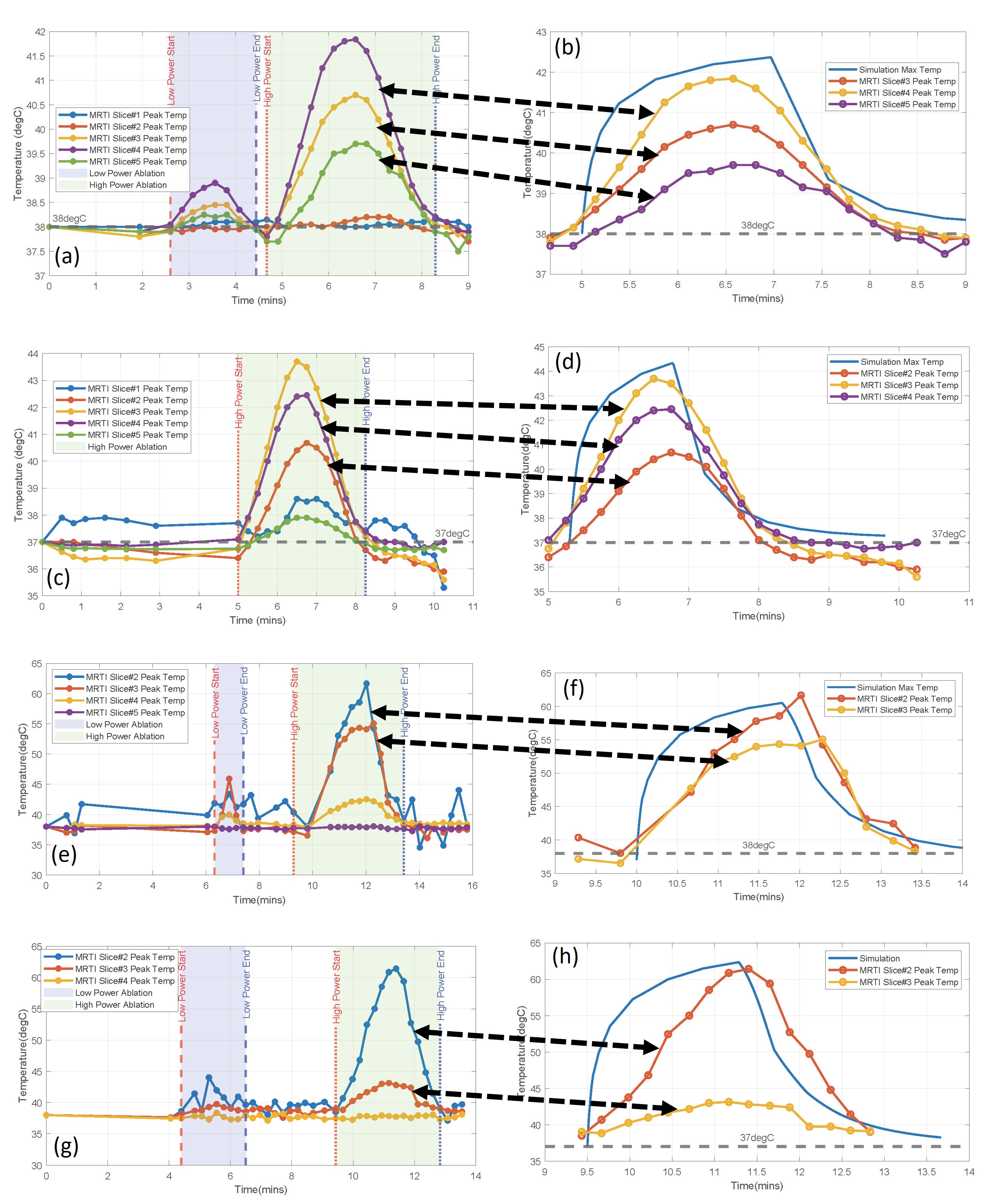}
    \caption{Multiple MRTI peak temperature results versus the simulation maximum temperature data using 360$^\circ$ probe. (a)-(b) Swine 1 experimental and simulation results. (c)-(d) Swine 2 data. (e)-(f) Swine 3 data. (g)-(h) Swine 5 data. For all the 360$^\circ$ probe data the simulation performed a relatively lower Pearson correlation coefficient (lowest at 0.7117), especially since the heating up duration temperature is not a rapid increase, and this may be caused by the brain tissue parameters misalignment between the experiment and simulation.}
    \label{fig:MRT360}
\end{figure*}

The rest of the lesions' results are also compared with the simulation data, and all the images can be found in Fig. \ref{fig:MRT180} and Fig. \ref{fig:MRT360}. The results of 180$^\circ$ is shown in Fig. \ref{fig:MRT180}, where the (a)-(b) shows the Swine 4 results with 180$^\circ$ probe under 3W acoustic power and 180s duration time condition, and the peak temperature reached 62$^\circ$C in experiment versus 63.2925$^\circ$C in simulation with Pearson correlation coefficient (PCC) of 0.9566. (c)-(d) shows the Swine 6 results with 180$^\circ$ probe under 4W acoustic power and 180s duration time condition, and the highest temperature reached 64$^\circ$C in experiment versus 64.9839$^\circ$C in simulation with PCC of 0.9275. Finally, the results of 360$^\circ$ are shown in Fig. \ref{fig:MRT360}, where the (a)-(b) shows the Swine 1 experimental and simulation results with 360$^\circ$ probe under 3W acoustic power and 90s duration time condition, and the highest temperature reached 41$^\circ$C in experiment versus 42.3701$^\circ$C in simulation with PCC of 0.9652. Note that the base temperature was 38$^\circ$C. (c)-(d) shows Swine 2 results with 360$^\circ$ probe under 3W acoustic power and 100s duration time condition, and the highest temperature reached 43$^\circ$C in experiment versus 44.3343$^\circ$C in simulation  with PCC of 0.9321. Note that there was no low-temperature testing process in this experiment. (e)-(f) Swine 3 results with 360$^\circ$ probe under 3W acoustic power and 120s duration time condition, and the highest temperature reached 61$^\circ$C in experiment versus 60.5230$^\circ$C in simulation with PCC of 0.8145. Note that the base temperature was also 38$^\circ$C in this experiment. (g)-(h) Swine 5 results with 360$^\circ$ probe under 4W acoustic power and 120s duration time condition, and the highest temperature reached 59$^\circ$C in experiment versus 61.1884$^\circ$C in simulation with PCC of 0.7117. Detailed results can be found in Table \ref{tbl:temp_volume_error}.

\subsection{Ablated CEM43 Map Results}

The previous section discussed the peak temperature change between MRTI data and simulation results, however, the CEM43 isodose is the desired indicator for evaluating the tissue damage in the field \cite{campwala2021predicting}. By using the volumetric MRTI measured the temperature changes at different distances from the focal point, and the necrotic zone boundaries were measured at CEM43\textgreater70 and multiplied the thickness of 5mm, we can calculate the ablated volume to compare with our simulation results.

\begin{figure}
    \centering
    \includegraphics[width=0.9\linewidth]{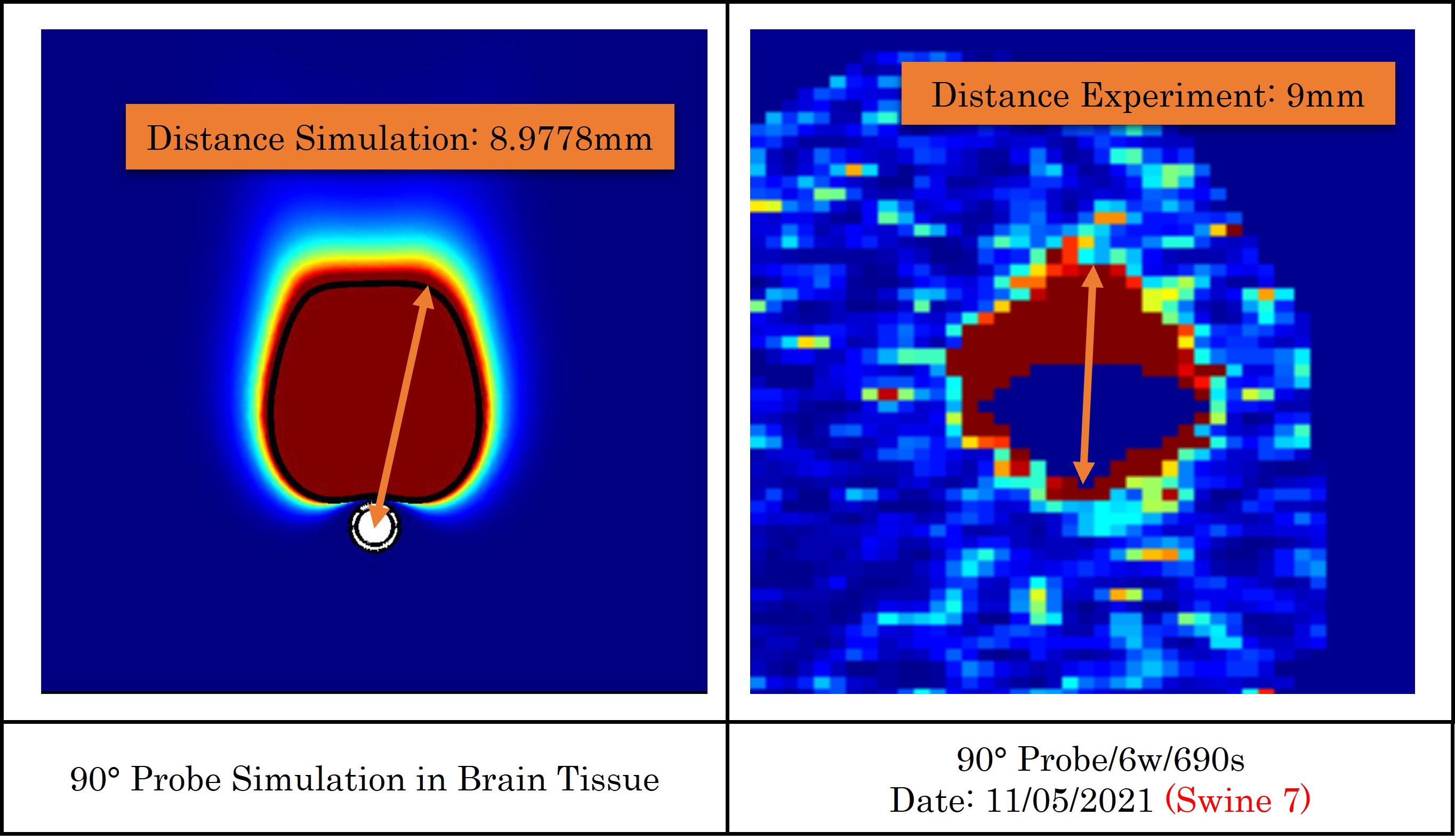}
    \caption{The simulation versus the experimental results of CEM43 of 70 isodose maps from 90$^\circ$ probes, and the largest distance from the center of the probe to the edge of the ablated area is 9mm in the experiment versus 8.9778 in simulation. }
    \label{fig:P90vs}
\end{figure}

The CEM43 of 70 isodose maps from 7 swine experiments are compared with the simulation results. An area was drawn around the margins of the ablated area with the lasso feature and then the area of the circled region in cm$^2$. Each ablated area was multiplied by the 5mm slice thickness if the ablation went through the posterior side of the slice. In slice 3 the largest distance from the probe center to the edge of the ablated area is 9mm and the ablated volume is 0.2344 cm$^3$, while in slice 3 the largest distance from the probe center to the edge of the ablated area is 7.4mm and the ablated volume is 0.2234 cm$^3$. Fig. \ref{fig:P90vs} shows the simulation versus the experimental results of the CEM43 of 70 isodose maps. The largest distance from the center of the probe to the edge of the ablated area is 9mm in the experiment versus 8.9778 in the simulation, which yields within 0.25\% error. Another metric, the Dice coefficient was used for evaluating the degree of overlap between simulation and experimental CEM43 isodose pattern. For swine 7, the Dice coefficient between simulation (center slice of transducer) and experimental results (MTRI slice 2 data) is 0.8496. 

\begin{figure}
    \centering
    \includegraphics[width=1\linewidth]{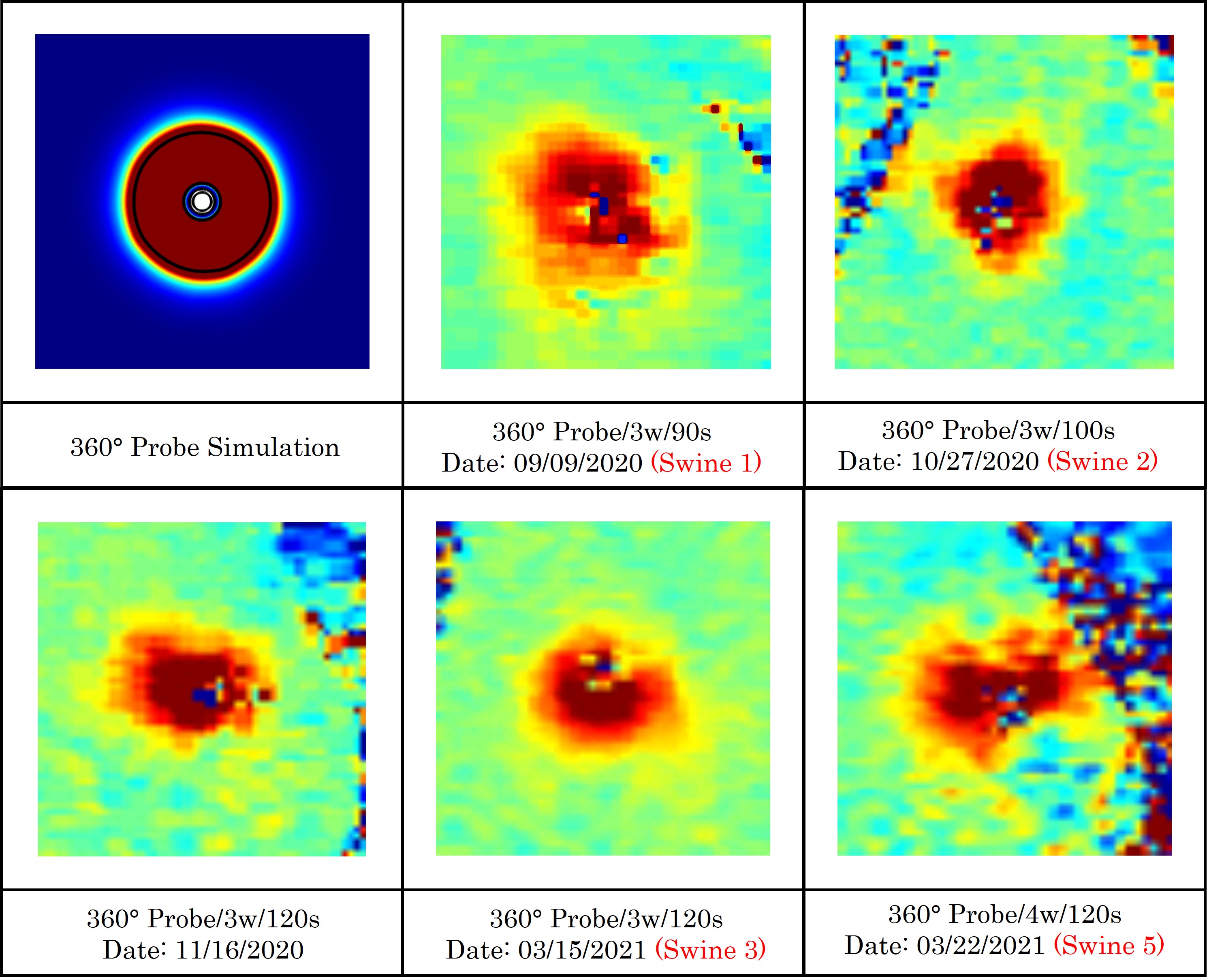}
    \caption{The simulation versus the experimental results of CEM43 of 70 isodose maps from 360$^\circ$ probe. Results follow the experimental data, which concludes all the \emph{in vivo} animal experiments the author had attended.}
    \label{fig:P360vs}
\end{figure}

\begin{figure}
    \centering
    \includegraphics[width=1\linewidth]{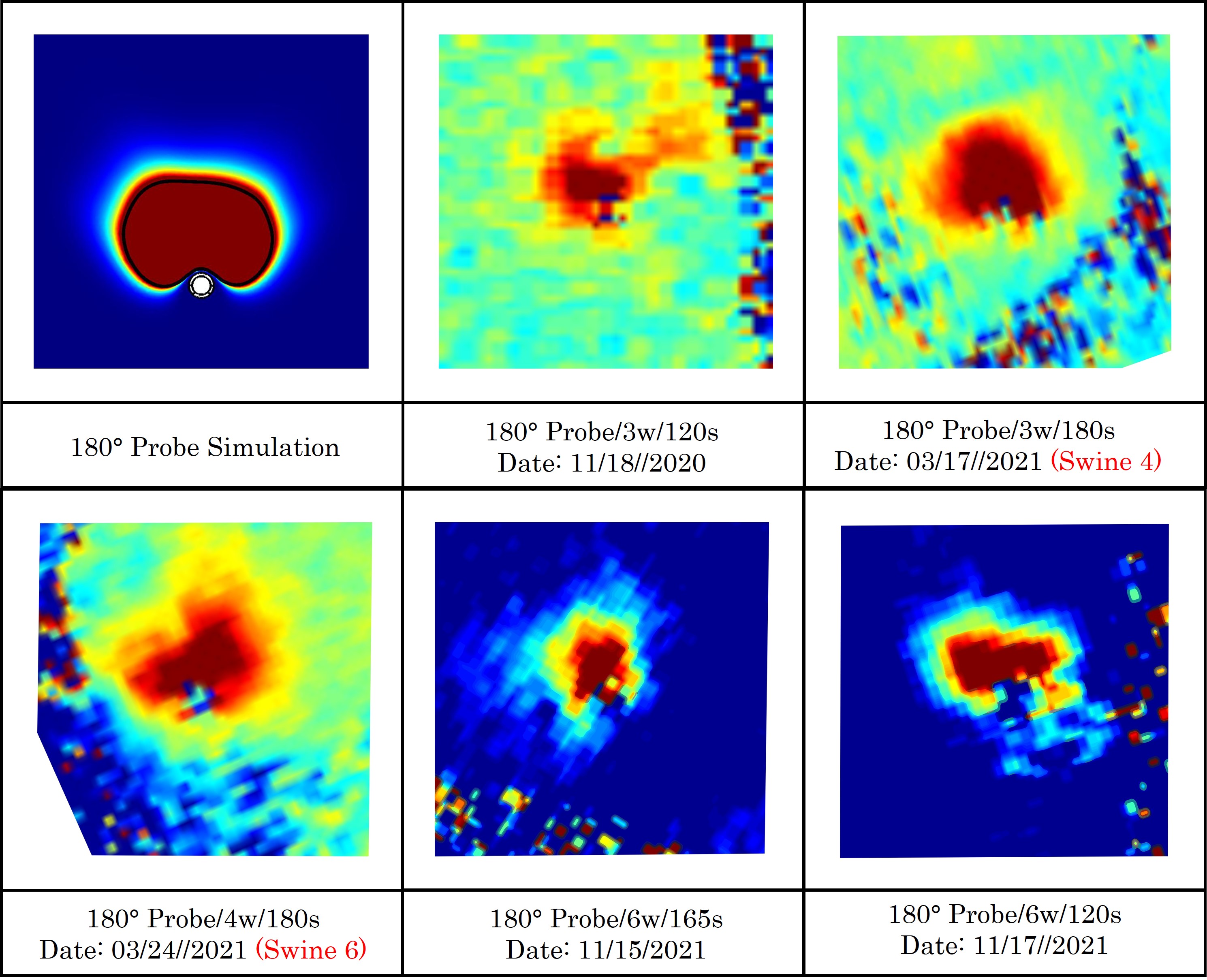}
    \caption{The simulation versus the experimental results of CEM43 of 70 isodose maps from 180$^\circ$ probe.Results follow the experimental data, which concludes all the \emph{in vivo} animal experiments the author had attended.}
    \label{fig:P180vs}
\end{figure}

Fig. \ref{fig:P360vs} and Fig. \ref{fig:P180vs} show the simulation versus the experimental results of CEM43 70 isodose maps using 360$^\circ$ and 180$^\circ$ probe results, and the Dice coefficient of selected swines experiment versus the simulation can be found in Table \ref{tbl:temp_volume_error}. The lowest Dice coefficient of experiment and simulation results by using 360$^\circ$ and 180$^\circ$ are 0.7021 and 0.8145 respectively. Note that these results contain all the \emph{in vivo} animal experiments the author had attended, which is beyond the 7 lesions mentioned in this study.
 
Detailed results of ablated volume data from the 7 swine experiments can be found in Table \ref{tbl:temp_volume_error}. By using the same method mentioned in MRTI, the ablated area calculated from simulated CEM43 of 70 isodose maps was multiplied by the 5mm slice thickness and summed together to get the ablated volume, which is also shown in Table \ref{tbl:temp_volume_error}. 

\begin{table*}[]
    \centering
    \caption{Highest Temperature and ablated volume compare and error between experimental and simulation results. *The Pearson correlation coefficient of experimental and simulation peak temperature curves. **The dice coefficient of the experimental  (largest ablated area slice) and simulation (center slice) with the ablated area under the same parameters.}
    \label{tbl:temp_volume_error}
    \resizebox{0.8\linewidth}{!}{
    \begin{tabular}{|c|c|c|c|c|c|c|c|c|}
    \hline
    Subject & \begin{tabular}[c]{@{}c@{}}MRTI\\ Max\\ Temp\\ ($^\circ$C)\end{tabular} & \begin{tabular}[c]{@{}c@{}}Predicted \\ Max\\ Temp \\ ($^\circ$C)\end{tabular} & \begin{tabular}[c]{@{}c@{}}Temp\\ Error\\ (\%)\end{tabular} & \begin{tabular}[c]{@{}c@{}}Pearson\\ Correlation\\ Coefficient*\end{tabular}& \begin{tabular}[c]{@{}c@{}}MRTI\\ Volume\\ (cm$^3$)\end{tabular} & \begin{tabular}[c]{@{}c@{}}Predicted\\ Volume\\ (cm$^3$)\end{tabular} & \begin{tabular}[c]{@{}c@{}}Volume\\ Error\\ (\%)\end{tabular} & \begin{tabular}[c]{@{}c@{}}Ablated \\ Area Dice\\ Coefficient**\end{tabular} \\ \hline
    
    Swine 1 & 41 & 42.3701 & 3.34 & 0.9652 & 0.4  & 0.4218 & 5.46 &0.8565\\ \hline
    Swine 2 & 43 & 44.3343 & 3.10 &0.9321 & 0.45  & 0.4646 & 3.25 &0.7316\\ \hline
    Swine 3 & 61 & 60.5230 & -0.78 & 0.8145 & 1  & 1.0110 & 1.10 &0.8152\\ \hline
    Swine 4 & 62 & 63.2925 & 2.08 & 0.9566 & 0.9 & 0.9375 & 4.17&0.8721\\ \hline
    Swine 5 & 59 & 61.1884 & 3.71 & 0.7117& 1.2 & 1.2677 & 5.64&0.8509\\ \hline
    Swine 6 & 64 & 64.9839 & 1.54 &0.9275 & 1.1 & 1.1631 & 5.74 &0.7021\\ \hline
    Swine 7 & 57 & 58.5823 & 2.78 &0.9527 & 0.46 & 0.4746 & 3.18 &0.8145\\ \hline
    \end{tabular}
    }
\end{table*}

\section{Conclusion}

In this work, we developed a 2D version simulation with real brain tissue parameters focused on thermal damage modeling, and validated with multiple \emph{in vivo} animal surgery. 

Taguchi-based sensitivity analysis was performed and was concluded that the voltage applied (V) and attenuation ($\alpha_{atten}$) are the most important parameters to determine the obtained maximum temperature in our bioheat transfer model. Consistent with other researchers' findings, we observed significant performance of ``voltage" contribution to ablation procedure \cite{jamil2013ranking}. We also suggested prioritizing attenuation ($\alpha_{atten}$) and thermal conductivity ($k$) and carefully considering these parameters in both simulation and experiments in bioheat transfer. 

The 2D thermal damage FEM simulation in brain tissue, including the validation with multiple \emph{in vivo} animal surgery. This model focused on real tissue thermal damage (CEM) instead of thermal propagation, which is what we desired in NBTU thermal ablation therapy. The highest temperature and ablated volume largest difference error between experimental and simulation results yield 2.1884$^\circ$C (3.71\%) and 0.0631cm$^3$ (5.74\%) respectively, with the lowest Pearson correlation coefficient (PCC) of peak temperature is 0.7117, and the lowest Dice coefficient of ablated area is 0.7021. These results show good agreement of accuracy between simulation and experiment. Note that the ones with higher temperature error (swine 5, 3.71\%) are not the same as the ones with higher volume error (swine 6, 5.74\%), this may be because the temperature error is more dependent on thermal conductivity and absorption effect, while the volume error is more dependent on thermal dissipation and perfusion rate.

\section{Discussion}

In this study, the changes of tissue properties are not considered, however as reported in \cite{prakash2012considerations} the perfusion and attenuation will change dynamically during the ablation procedure, especially the ablated section will be different compared to un-ablated status both in healthy tissue and tumor cells. Both low PCC and Dice coefficient were performed when using 360$^\circ$, especially the heating up duration and some unexpected edge of ablation. Considering the sensitivity analysis we performed, the attenuation is ranking 2$^{nd}$ among parameters in our simulation, and it can change up to $\pm$ 50\% during the ablation procedure in soft tissue \cite{prakash2012considerations}, which will affect the ablation performance significantly. Also, the surrounding vasculature system may further induce these unexpected misalignments of ablation boundaries. Moreover, the heat transfer method between the above two different status tissues should also be considered to make the modeling more accurate. Future work may focus on dynamic parameters integrated simulation, especially considering the value shift from attenuation, and add more physics nodes with heat transfer among blood vascular system. The current sensitivity study only focuses on multiple parameters. In the future, we may consider involving some interactions of parameters in the analysis to get a more comprehensive ranking of all possible factors affecting the temperature and damage isodose.

\bibliographystyle{IEEEtran}
\bibliography{Main}

\end{document}